\documentclass[11pt]{article}

\usepackage[T1]{fontenc}
\usepackage[utf8]{inputenc}
\usepackage[english]{babel}
\usepackage{lmodern}
\usepackage{microtype}
\usepackage[a4paper,margin=27mm]{geometry}
\usepackage{amsmath,amssymb,amsthm,mathtools}
\usepackage{booktabs}
\usepackage{enumitem}
\usepackage{xcolor}
\usepackage{hyperref}
\usepackage[numbers,sort&compress]{natbib}

\hypersetup{
  colorlinks=true,
  linkcolor=blue,
  citecolor=blue,
  urlcolor=blue,
  pdftitle={Algorithmic Idealism I: Reconceptualizing Reality Through Information and Experience},
  pdfauthor={Krzysztof Sienicki}
}

\newtheorem{postulate}{Postulate}

\newcommand{\M}{\mathbf{M}}
\newcommand{\SU}{\mathcal{S}}
\newcommand{\bits}{\{0,1\}^{*}}

\title{\textbf{Algorithmic Idealism I: Reconceptualizing Reality Through Information and Experience}}
\author{Krzysztof Sienicki\thanks{Chair of Theoretical Physics of Naturally Intelligent Systems, Lipowa 2/Topolowa 19, 05-807 Podkowa Le\'sna, Poland, European Union.}}
\date{Revised August 2026}

\begin{document}

\maketitle

\begin{abstract}
Algorithmic Idealism is a first-person approach to foundations in which momentary ``self states'' are treated as fundamental and a universal method of induction supplies objective likelihoods for future self states. The framework, developed by Markus P.~M\"uller, is motivated in part by quantum theory but is not itself an interpretation of quantum mechanics. Its current form is best understood as a conceptual framework admitting several mathematical realizations. A preliminary realization---the bit model---uses algorithmic probability and admits rigorous results concerning asymptotic agreement with computable stochastic environments and the emergence of an effective external world. More general claims about quantum phenomena, Boltzmann brains, simulations, duplication, and personal identity require additional qualifications.

This article revises an earlier 2024 assessment in light of M\"uller's peer-reviewed formulation in \emph{Foundations of Physics}, the 2026 analysis of ``Restriction A'' by Jones and M\"uller, and the 2026 adversarial collaboration between McQueen and M\"uller. We distinguish (i) the postulates of Algorithmic Idealism, (ii) theorems established for a particular formal model, (iii) proposed physical predictions, and (iv) philosophical extrapolations. In particular, algorithmic transition probabilities are not identified by definition with Born probabilities, informational duplication does not by itself settle numerical personal identity, and the Boltzmann-brain and simulation claims remain theory-dependent predictions rather than generally established consequences of physics. The resulting picture is more modest than the original presentation, but also more scientifically precise: Algorithmic Idealism is a mathematically motivated research program whose strongest results are model-specific and whose broader empirical and metaphysical status remains open.
\end{abstract}

\tableofcontents

\section{Introduction}

Algorithmic Idealism asks a question that is deliberately different from the traditional foundational question ``What is the case in the world?'' Its guiding question is instead: \emph{What should an agent expect to experience next?} In M\"uller's formulation, the aim is to make this question mathematically precise even in situations in which ordinary third-person physical descriptions do not straightforwardly determine first-person expectations \cite{Mueller2026AI}.

This perspective is motivated by several families of problems. Quantum theory supplies probabilities for future measurement outcomes, while extended Wigner's-friend scenarios expose difficulties in combining all agents' observations into a single joint probabilistic description under additional assumptions such as locality and no-superdeterminism \cite{Bong2020,JonesMueller2026}. Cosmology raises self-locating problems such as the Boltzmann-brain puzzle \cite{Carroll2020,Page2008}. Duplication and teletransportation scenarios raise questions about first-person continuation that are not exhausted by a third-person inventory of physical copies \cite{Parfit1984}. The simulation argument similarly invites a distinction between the number of physical or computational realizations and the probabilities an observer should assign to future experience \cite{Bostrom2003}.

The 2024 version of this article described Algorithmic Idealism too strongly in several places. It sometimes treated a proposed resolution as an established result, identified algorithmic probabilities too directly with quantum probabilities, and inferred claims about personal identity that do not follow from the formal postulates. The purpose of the present revision is therefore not merely stylistic. It is to separate four logically different levels:

\begin{enumerate}[label=(\roman*)]
  \item the general postulates of Algorithmic Idealism;
  \item rigorous results in particular mathematical realizations, especially the bit model;
  \item physical or cosmological predictions suggested by the framework;
  \item philosophical interpretations and extrapolations that are not mathematical theorems.
\end{enumerate}

This separation is particularly important in 2026. M\"uller's main conceptual paper is now peer reviewed and published \cite{Mueller2026AI}; the structural motivation from Wigner's-friend and duplication scenarios has been developed into the notion of \emph{Restriction A} \cite{JonesMueller2026}; and the coherence and explanatory force of the framework have become the subject of an explicit realist--idealist adversarial collaboration \cite{McQueenMueller2026}. Thus Algorithmic Idealism should now be assessed as an active research program rather than as a settled solution to a list of foundational problems.

\section{The Framework}

\subsection{Restriction A as motivation}

Jones and M\"uller introduce the expression \emph{Restriction A} for a structural limitation of physical theories: in some scenarios, a theory cannot supply a single joint probabilistic description of the observations of all agents \cite{JonesMueller2026}. In extended Wigner's-friend scenarios this issue can arise in quantum theory under assumptions such as locality and no-superdeterminism. Importantly, Jones and M\"uller also construct classical duplication scenarios displaying an analogous structural difficulty. The point is therefore broader than the quantum measurement problem.

Restriction A must be formulated carefully. It does not say that physics generally fails to predict observations. In ordinary laboratory situations, physical theories provide extremely successful and intersubjectively testable probabilities. Rather, the claim is that there are unusual situations in which the future observations relevant to different agents cannot all be represented as marginals of one common probability distribution without giving up some background assumptions. For a single agent, related duplication or simulation scenarios may also leave ordinary third-person physics without a unique rule for the agent's own first-person credences.

Algorithmic Idealism can be viewed as one proposed response to the single-agent version of this problem. Instead of deriving every first-person probability from an assumed external world, it postulates a direct law on momentary informational states.

\subsection{Self states}

M\"uller's first postulate can be summarized as follows \cite{Mueller2026AI}.

\begin{postulate}[Self States]
There is a set $\SU$ of self states: abstract patterns that can characterize a system at a given subjective moment. Everything that the theory says about the system's past and future is conditioned on its current self state.
\end{postulate}

The term \emph{self state} should not be read as a complete neuroscientific theory of a person, nor as a primitive notion of consciousness. In the formal framework, self states are abstract mathematical objects. Only some of them may admit an interpretation as momentary states of human or artificial agents. M\"uller deliberately avoids making ``agent'', ``observer'', ``consciousness'', or ``belief'' primitive notions of the theory \cite{Mueller2026AI}.

A central and counterintuitive part of the postulate is \emph{unembeddedness}: the fundamental description of the current self state does not include a primitive spatiotemporal address in an independently given external world. A familiar external world is instead intended to emerge as an effective description under suitable conditions. This is a substantive hypothesis of the framework, not a consequence of algorithmic information theory alone.

\subsection{State change}

The second postulate concerns the transition from a current self state $x$ to a future self state $y$.

\begin{postulate}[State Change]
Given a current self state $x\in\SU$, there is an objective likelihood structure for future self states $y\in\SU$. This likelihood is supplied by a universal method $M$ of induction.
\end{postulate}

The word ``likelihood'' is intentionally broader than a single ordinary probability distribution. Different mathematical realizations may use a probability distribution, a family of distributions, imprecise probabilities, or another formally specified likelihood structure \cite{Mueller2026AI}. Consequently, Algorithmic Idealism is better described as a \emph{framework} than as a unique completed mathematical theory.

The crucial scientific hypothesis is the extrapolation of induction: a universal inductive rule is assumed to remain applicable not only in ordinary situations where we already appear embedded in a stable world, but also in exotic scenarios involving duplication, simulations, severe self-location ambiguity, or other cases in which ordinary physical reasoning does not by itself fix first-person probabilities.

\section{The Bit Model and What Is Actually Proved}

\subsection{Preliminary realization}

The technically best-developed realization predates the name ``Algorithmic Idealism''. In M\"uller's 2020 paper, observer states are finite binary strings and universal induction determines the probabilities for their continuation \cite{Mueller2020}. In the terminology of the later framework, this becomes the \emph{bit model}.

The self-state space is
\begin{equation}
\SU=\bits.
\end{equation}
If the current state is $x$, the next state has the form $xb$, where $b\in\{0,1\}$. For a universal monotone Turing machine $U$, one introduces a universal semimeasure $\M_U$. Schematically,
\begin{equation}
\M_U(x)
=\sum_{p:\,U(p)=x*}2^{-|p|},
\end{equation}
where $U(p)=x*$ indicates that the output generated by program $p$ begins with $x$. Conditional continuation weights are then obtained from
\begin{equation}
\M_U(b\mid x)
=\frac{\M_U(xb)}{\M_U(x)},
\end{equation}
whenever the denominator is nonzero. The exact framework is slightly more subtle than simply selecting one preferred universal machine: M\"uller's formulation treats predictions as those features that are sufficiently invariant under the admissible choices of universal machine \cite{Mueller2020,Mueller2026AI}.

This is materially different from the oversimplified statement
\begin{equation}
P(y\mid x)\propto 2^{-K(y\mid x)}.
\end{equation}
Kolmogorov complexity is closely related to algorithmic probability, but the bit-model transition rule is formulated using universal semimeasures and their conditional continuations, not by assigning an arbitrary future self state a probability proportional to a single conditional complexity.

\subsection{Solomonoff convergence}

One rigorous reason universal induction is relevant is the convergence theorem for computable stochastic sources. If observations are generated by a computable probability measure $\mu$, then Solomonoff prediction converges to the corresponding conditional probabilities under standard assumptions. In the binary notation relevant here, one obtains an almost-sure asymptotic statement of the form
\begin{equation}
\lim_{n\to\infty}
\left|
\M_U(b\mid x_1\ldots x_n)
-
\mu(b\mid x_1\ldots x_n)
\right|=0,
\qquad b\in\{0,1\},
\end{equation}
for $\mu$-typical sequences \cite{Solomonoff1964,LiVitanyi,Mueller2020}.

This result is important, but its interpretation must be kept precise. It says that universal induction learns computable probabilistic regularities asymptotically. It does \emph{not} by itself prove that the physical world is generated by a computable measure, that a particular self-state representation is correct, or that the Born rule is identical to algorithmic probability.

\subsection{Emergent external-world descriptions}

M\"uller's 2020 analysis contains rigorous results showing that, in the bit model and under specified assumptions, an observer's data can asymptotically behave as if embedded in an external computable probabilistic process \cite{Mueller2020}. This is one of the framework's strongest mathematical achievements. It makes precise a limited sense in which an ``external world'' can be an emergent effective description rather than a fundamental postulate.

The qualification \emph{in the bit model} is essential. The bit model assumes that self states grow by appending information and therefore do not fundamentally forget. M\"uller now explicitly treats this ``erasure problem'' as a limitation and is seeking more general formulations that allow genuine information loss while retaining analogues of the existing theorems \cite{Mueller2026AI}. Accordingly, it is premature to state without qualification that Algorithmic Idealism in general has proved the emergence of the complete physical world we observe.

\section{Quantum Theory: Motivation, Not Identification}

\subsection{Not an interpretation of quantum mechanics}

A major correction to the earlier version of this article is required here. Algorithmic Idealism is \emph{not} intended as a new interpretation of quantum mechanics in the ordinary sense. M\"uller now states this explicitly: quantum theory motivates the first-person question and provides an important empirical ``litmus test'', but Algorithmic Idealism is a separate framework intended to make additional predictions in regimes not covered by ordinary quantum theory \cite{Mueller2026AI}.

This distinction prevents a conceptual conflation. Interpretations such as QBism or Berghofer's degrees-of-epistemic-justification interpretation (DEJI) concern the meaning and normative role of the quantum state and Born probabilities \cite{FuchsMerminSchack2014,Berghofer2024}. Algorithmic Idealism instead proposes a law governing transitions between self states. The two projects may be philosophically sympathetic, but they are not the same theory.

\subsection{Born probabilities versus algorithmic probabilities}

The Born rule gives, for a quantum state $\rho$ and a POVM $\{E_i\}$,
\begin{equation}
P(i)=\operatorname{Tr}(\rho E_i).
\end{equation}
These operational probabilities are empirically confirmed predictions of quantum theory. By contrast, the bit model uses conditional algorithmic probabilities $\M_U(b\mid x)$ on binary self-state continuations.

There is no general identity
\begin{equation}
\M_U(b\mid x)
=\operatorname{Tr}(\rho E_b)
\end{equation}
by definition. For Algorithmic Idealism to reproduce quantum experiments, its first-person predictions must agree with Born-rule statistics in the appropriate operational regime, either exactly or to controlled approximation. M\"uller's 2020 work contains preliminary results suggesting that nonclassical phenomena such as Bell-inequality violation together with no-signalling can arise in special constructions, but these results depend on substantial assumptions and do not amount to a derivation of quantum theory in full \cite{Mueller2020}.

Thus the scientifically defensible statement is conditional:

\begin{quote}
In regimes where a realization of Algorithmic Idealism successfully reproduces ordinary quantum predictions, its induced first-person probabilities must agree with the relevant Born-rule probabilities for observable outcomes. They should not be identified with Born probabilities by definition.
\end{quote}

\subsection{The measurement problem}

The framework may motivate a first-person reformulation of measurement, but it should not be advertised as having solved the quantum measurement problem. The measurement problem is a family of questions concerning the relation among unitary dynamics, definite outcomes, state update, and ontology. Algorithmic Idealism changes the foundational target by taking future self states rather than an observer-independent wavefunction history as fundamental. That may dissolve or bypass some versions of the problem, but whether this counts as a solution depends on what one demands of a solution and on whether the framework can recover the full empirical and mathematical structure of quantum theory.

The 2026 work of Jones and M\"uller reinforces this caution: the structural difficulty highlighted by Wigner's-friend scenarios is argued to extend beyond quantum mechanics, including to classical duplication scenarios \cite{JonesMueller2026}. It is therefore misleading to present Algorithmic Idealism merely as a special answer to wavefunction collapse.

\section{Boltzmann Brains, Simulations, and Duplication}

\subsection{Boltzmann brains}

A Boltzmann-brain problem arises in cosmological models in which random fluctuations generate observers with apparently coherent but false memories, potentially in numbers vastly exceeding ordinary evolved observers \cite{Page2008,Carroll2020}. The standard inferential step from ``there are many more Boltzmann brains'' to ``I should expect to be a Boltzmann brain'' requires an additional self-locating or typicality principle; it is not supplied by the cosmological dynamics alone \cite{JonesMueller2026}.

Algorithmic Idealism rejects the idea that a current self state must fundamentally be one particular physical realization whose identity is unknown. Instead, first-person continuation is governed directly by the current self state and the inductive law. In the bit model, this leads to a result according to which the numerical abundance of Boltzmann-brain realizations does not by itself dominate the observer's future first-person probabilities \cite{Mueller2020}.

This is a genuine and distinctive proposal. Nevertheless, the strongest wording should remain theory-relative. It is better to say that Algorithmic Idealism \emph{predicts} that an agent should not bet on a Boltzmann-brain continuation merely because Boltzmann-brain realizations are numerically dominant. It is too strong to claim that the Boltzmann-brain problem has been unconditionally ``eliminated'' for cosmology as a whole. Indeed, the 2026 adversarial collaboration identifies precisely the coherence of Algorithmic Idealism and its ability to resolve observer paradoxes such as the Boltzmann-brain problem as an active point of disagreement \cite{McQueenMueller2026}.

\subsection{The simulation hypothesis}

Bostrom's simulation argument concerns a probabilistic inference from assumptions about technologically mature civilizations and the number of simulated observers \cite{Bostrom2003}. Algorithmic Idealism approaches the issue differently. The number of physical or computational realizations of a self state is not, by itself, the quantity that fixes first-person continuation probabilities.

This does not imply the simple slogan that simulated and non-simulated existence are ``identical''. Rather, the framework claims that a self state is fundamentally unembedded and that an effective embedding into a particular external process is secondary. In some circumstances, a simulation may track the effective first-person dynamics accurately; in others, external interventions or termination may cause the physical simulation and the first-person transition law to diverge. M\"uller consequently argues that shutting down a simulation does not \emph{generally} imply termination of its inhabitants in the first-person sense \cite{Mueller2026AI}. This is a counterintuitive theory-dependent prediction, not a conclusion of computer science or physics independently of Algorithmic Idealism.

\subsection{Teletransportation and personal identity}

Parfit's teletransportation scenarios separate questions about bodily continuity, psychological continuity, duplication, and numerical identity \cite{Parfit1984}. Algorithmic Idealism provides a rule for first-person transition likelihoods from a current self state in such cases. It also treats multiple physical realizations of the same present self state equivalently at the fundamental level.

What does \emph{not} follow is the metaphysical theorem
\begin{equation}
\text{informational identity}
\quad\Longrightarrow\quad
\text{numerical personal identity}.
\end{equation}
The framework does not need to posit a persisting metaphysical person as a primitive at all. It predicts transitions among momentary states. Accordingly, claims such as ``the uploaded copy is literally the same person'' go beyond the formalism unless an additional theory of personal identity is supplied.

A more accurate conclusion is:

\begin{quote}
Algorithmic Idealism treats physically distinct realizations of the same present self state as equivalent for the purpose of determining first-person transition likelihoods. This may reduce the theoretical role of an ``original versus copy'' distinction, but it does not by itself settle the metaphysics of numerical personal identity.
\end{quote}

\section{Philosophical Implications}

\subsection{Ontological economy and emergent worlds}

Algorithmic Idealism is ontologically revisionary but should not be described merely as the denial that the external world is real. Its central proposal is instead that external-world structure is not fundamental. In standard regimes, an agent may be \emph{effectively embedded} in a stable, simple, probabilistic world, and ordinary physical explanations remain valid at that emergent level \cite{Mueller2020,Mueller2026AI}.

The distinction between \emph{fundamental} and \emph{effective} is essential. A theory can deny that spacetime location is a primitive property of self states while still recovering an excellent mechanistic description in terms of brains, laboratories, planets, and physical laws whenever those structures supply the best stable predictive model of the data.

\subsection{Intersubjectivity}

The earlier manuscript contained a tension: it described intersubjectivity both as a central prediction and as an unresolved problem. The corrected position is more nuanced. In the 2020 bit model, M\"uller proved results supporting an emergent relation between first-person and third-person descriptions under appropriate conditions \cite{Mueller2020}. This demonstrates that a first-person theory need not collapse into simple solipsism.

However, the generalization to richer self-state models, especially models allowing erasure, is not complete. Moreover, Restriction A explicitly warns that there need not be a single joint description of all agents' observations in every conceivable experiment \cite{JonesMueller2026}. Hence intersubjective objective reality is best viewed as an approximate and domain-dependent emergent structure rather than an unconditional universal feature.

\subsection{Ethics}

Algorithmic Idealism raises ethically important questions about simulations, duplication, digital reconstruction, and artificial agents, but it does not by itself supply a moral theory. A transition law tells us which future self states are likely; it does not directly assign moral status, rights, welfare, or responsibility.

In particular, one should not infer that informationally identical states necessarily have identical moral status in every context, nor that termination of a physical implementation is automatically morally harmless because some first-person continuation may remain possible according to the theory. Ethical conclusions require additional premises about consciousness, welfare, identity, and moral patienthood. The framework is therefore best regarded as generating new ethical questions rather than solving them.

\section{Strengths, Limitations, and Falsifiability}

\subsection{Strengths}

The framework has several genuine strengths.

First, it isolates a precise target that ordinary third-person physics does not always obviously answer: objective first-person likelihoods in exotic scenarios. Second, it connects this target to a mathematically mature theory of universal induction. Third, the 2020 bit model proves nontrivial emergence results rather than merely asserting that an external world appears. Fourth, the 2026 work on Restriction A gives the motivation a broader structural basis that is not restricted to the quantum measurement problem \cite{JonesMueller2026}. Finally, the framework makes unusual predictions in regimes such as duplication, Boltzmann brains, and simulation, which creates at least the possibility of empirical or conceptual discrimination from competing approaches.

\subsection{Model dependence and the erasure problem}

The principal mathematical limitation is that the strongest theorems presently apply to a preliminary model. In the bit model, the state retains its complete prefix history. Human observers plainly forget, compress, overwrite, and reconstruct information. M\"uller therefore regards a formulation with genuine erasure as an open problem \cite{Mueller2026AI}.

This matters because conclusions proved for the bit model cannot automatically be promoted to the whole conceptual framework. A successful generalization should specify a new state space $\SU$, a precise universal inductive structure $M$, and analogues of the convergence and emergence results.

\subsection{Machine dependence and noncomputability}

Algorithmic information quantities depend on a choice of universal description language up to invariance theorems. Exact Kolmogorov complexity and Solomonoff's universal semimeasure are also noncomputable in general \cite{LiVitanyi}. This does not invalidate their use in an in-principle physical theory, but it complicates claims of practical calculability and unique finite predictions.

M\"uller's current framework addresses universal-machine dependence by focusing on predictions that are invariant, or sufficiently stable, across universal descriptions \cite{Mueller2020,Mueller2026AI}. Whether this strategy yields sufficiently sharp predictions in richer models remains an important technical question.

\subsection{Empirical status}

The statement that Algorithmic Idealism is ``untestable'' is too strong, but so is the claim that it is already well confirmed. Its standard-regime motivation is partly indirect: universal induction has convergence properties, and the bit model can recover effective external-world behavior under assumptions. Its most distinctive predictions concern unusual experiments whose outcomes may be privately rather than intersubjectively verifiable.

M\"uller argues that a single theory can be supported by its successful ordinary predictions and then trusted in exotic private regimes \cite{Mueller2026AI}. A critic may reasonably object that privately verifiable predictions are unusually difficult to incorporate into communal scientific testing. This tension is not a logical contradiction, but it is a central methodological issue for the research program.

There are also possible routes to falsification in principle. If every plausible rigorous realization predicts phenomenology incompatible with established physics, if no erasure-capable model reproduces the bit-model emergence results, or if the framework cannot reproduce robust quantum statistics in standard experiments, its central ambitions would fail. Conversely, a successful derivation of nonclassical quantum structure from the universal first-person law would substantially strengthen the program.

\section{The 2026 Realism--Idealism Adversarial Collaboration}

The most important update since the original version of this article is the adversarial collaboration between Kelvin J.~McQueen and Markus P.~M\"uller \cite{McQueenMueller2026}. The two authors deliberately begin from opposed positions. McQueen defends a realist picture in which an external physical world is fundamental; M\"uller defends the view that first-person states are fundamental and shared physical worlds are emergent.

The significance of this work is methodological as well as substantive. It shows that, by mid-2026, the core dispute is no longer well described as ``Algorithmic Idealism has solved the paradoxes and only details remain.'' Instead, the disagreement concerns the coherence of the formal model itself and whether the alleged resolutions of observer paradoxes---especially the Boltzmann-brain problem---are genuinely explanatory.

This is a healthy development for the research program. It sharpens the burden of proof. A mature assessment should ask at least four questions:

\begin{enumerate}[label=(\arabic*)]
  \item Can a general, erasure-capable model be formulated with the required mathematical rigor?
  \item Does it recover ordinary physical predictions, including quantum statistics, without importing them by hand?
  \item Are its exotic first-person predictions invariant enough under representational choices to be physically meaningful?
  \item Do those predictions solve the intended paradoxes, or merely replace familiar self-location assumptions with a different primitive law?
\end{enumerate}

The adversarial collaboration does not settle these questions. It establishes that they are the correct questions on which future evaluation should focus.

\section{Conclusions}

Algorithmic Idealism remains one of the more unusual contemporary attempts to formulate a first-person law of nature. Its strongest contribution is not a proof that reality ``is information'', nor a completed solution to the quantum measurement problem, personal identity, or cosmological self-location. Its stronger and more defensible contribution is methodological: it identifies a class of first-person prediction problems, proposes universal induction as a lawlike response, and demonstrates in a preliminary mathematical model that an apparently objective, law-governed world can emerge asymptotically from a first-person starting point.

The distinction between the general framework and the bit model is therefore essential. Rigorous theorems exist, but they are model-specific. The framework's broader claims about quantum theory, simulations, Boltzmann brains, and duplication remain conditional on successful generalization and on the physical interpretation of its transition law.

The revised assessment is consequently more restrained than the 2024 version. Algorithmic Idealism should not currently be presented as a unified established solution to foundational physics and metaphysics. It is better described as a mathematically informed research program with several rigorous preliminary results, distinctive proposed predictions, and substantial unresolved questions. The peer-reviewed 2026 formulation, the development of Restriction A, and the realism--idealism adversarial collaboration have made the program scientifically clearer precisely by making its open problems more explicit.

\section*{Acknowledgments}

I am grateful to Markus P.~M\"uller from the Institute for Quantum Optics and Quantum Information (IQOQI), Vienna, for helpful comments on an earlier version of this work.


\begin{thebibliography}{99}

\bibitem{Mueller2026AI}
M.~P. M\"uller,
``Algorithmic Idealism: What Should You Believe to Experience Next?,''
\emph{Foundations of Physics} \textbf{56}, 11 (2026).
\href{https://doi.org/10.1007/s10701-026-00913-1}{doi:10.1007/s10701-026-00913-1};
\href{https://arxiv.org/abs/2412.02826}{arXiv:2412.02826}.

\bibitem{Mueller2020}
M.~P. M\"uller,
``Law without law: from observer states to physics via algorithmic information theory,''
\emph{Quantum} \textbf{4}, 301 (2020).
\href{https://doi.org/10.22331/q-2020-07-20-301}{doi:10.22331/q-2020-07-20-301};
\href{https://arxiv.org/abs/1712.01826}{arXiv:1712.01826}.

\bibitem{JonesMueller2026}
C.~L. Jones and M.~P. M\"uller,
``On the significance of Wigner's Friend in contexts beyond quantum foundations,''
\emph{Quantum} \textbf{10}, 2147 (2026).
\href{https://doi.org/10.22331/q-2026-06-30-2147}{doi:10.22331/q-2026-06-30-2147};
\href{https://arxiv.org/abs/2402.08727}{arXiv:2402.08727}.

\bibitem{McQueenMueller2026}
K.~J. McQueen and M.~P. M\"uller,
``Realism about the external world: an adversarial collaboration,''
\href{https://arxiv.org/abs/2607.16379}{arXiv:2607.16379} (2026).

\bibitem{Bong2020}
K.-W. Bong, A. Utreras-Alarc\'on, F. Ghafari, Y.-C. Liang, N. Tischler,
E.~G. Cavalcanti, G.~J. Pryde, and H.~M. Wiseman,
``A strong no-go theorem on the Wigner's friend paradox,''
\emph{Nature Physics} \textbf{16}, 1199--1205 (2020).
\href{https://doi.org/10.1038/s41567-020-0990-x}{doi:10.1038/s41567-020-0990-x}.

\bibitem{Bell1964}
J.~S. Bell,
``On the Einstein Podolsky Rosen paradox,''
\emph{Physics Physique Fizika} \textbf{1}, 195--200 (1964).
\href{https://doi.org/10.1103/PhysicsPhysiqueFizika.1.195}{doi:10.1103/PhysicsPhysiqueFizika.1.195}.

\bibitem{Berghofer2024}
P. Berghofer,
``Quantum Probabilities Are Objective Degrees of Epistemic Justification,''
\href{https://arxiv.org/abs/2410.19175}{arXiv:2410.19175} (2024).

\bibitem{FuchsMerminSchack2014}
C.~A. Fuchs, N.~D. Mermin, and R. Schack,
``An introduction to QBism with an application to the locality of quantum mechanics,''
\emph{American Journal of Physics} \textbf{82}, 749--754 (2014).
\href{https://doi.org/10.1119/1.4874855}{doi:10.1119/1.4874855}.

\bibitem{Solomonoff1964}
R.~J. Solomonoff,
``A formal theory of inductive inference. Part I,''
\emph{Information and Control} \textbf{7}, 1--22 (1964).
\href{https://doi.org/10.1016/S0019-9958(64)90223-2}{doi:10.1016/S0019-9958(64)90223-2}.

\bibitem{LiVitanyi}
M. Li and P. Vit\'anyi,
\emph{An Introduction to Kolmogorov Complexity and Its Applications},
4th ed., Springer, Cham (2019).
\href{https://doi.org/10.1007/978-3-030-11298-1}{doi:10.1007/978-3-030-11298-1}.

\bibitem{Parfit1984}
D. Parfit,
\emph{Reasons and Persons},
Oxford University Press, Oxford (1984).

\bibitem{Bostrom2003}
N. Bostrom,
``Are You Living in a Computer Simulation?,''
\emph{The Philosophical Quarterly} \textbf{53}, 243--255 (2003).
\href{https://doi.org/10.1111/1467-9213.00309}{doi:10.1111/1467-9213.00309}.

\bibitem{Page2008}
D.~N. Page,
``Return of the Boltzmann brains,''
\emph{Physical Review D} \textbf{78}, 063536 (2008).
\href{https://doi.org/10.1103/PhysRevD.78.063536}{doi:10.1103/PhysRevD.78.063536}.

\bibitem{Carroll2020}
S.~M. Carroll,
``Why Boltzmann brains are bad,''
in \emph{Current Controversies in Philosophy of Science},
Routledge (2020), pp.~7--20.
\href{https://arxiv.org/abs/1702.00850}{arXiv:1702.00850}.

\end{thebibliography}
\end{document}